\documentclass[a4paper]{llncs2e/llncs}

\usepackage[english]{babel}
\usepackage{csquotes}
\usepackage{amsmath,amssymb}
\usepackage{graphicx}
\usepackage{enumerate}
\usepackage{verbatim}

\usepackage{cite}

\usepackage[usenames,dvipsnames]{xcolor}
\usepackage{mathtools}
\usepackage{bm}
\usepackage{subcaption}
\usepackage[hidelinks]{hyperref}
\usepackage{stmaryrd} 

\newcommand{\env}[2]{\begin{#1}#2\end{#1}}%
\newcommand{\envparam}[3][]{\begin{#2}[#1]#3\end{#2}}%

\spnewtheorem*{theorem*}{Theorem}{\bf}{\it}
\spnewtheorem*{lemma*}{Lemma}{\bf}{\it}
\spnewtheorem*{definition*}{Definition}{\bf}{\it}
\spnewtheorem*{conjecture*}{Conjecture}{\it}{}

\def \N {\mathbb{N}}
\def \Z {\mathbb{Z}}
\def \Q {\mathbb{Q}}

\def \C {\mathbb{C}}

\def \A {\mathcal{A}}

\def \u {{\bm{u}}}
\def \v {{\bm{v}}}
\def \w {{\bm{w}}}

\DeclarePairedDelimiter\abs{\lvert}{\rvert}
\DeclareMathOperator{\supp}{supp}
\newcommand{\makeinner}[2]{ \, #1 \mid #2 \, }
\newcommand{\makeset}[2]{\{ \makeinner{#1}{#2} \}}
\newcommand{\makesetbig}[2]{\big\{ \, #1 \; \big| \; #2 \, \big\}}%
\newcommand{\restr}[1]{\!\!\upharpoonright_{ #1 }}
\newcommand{\uu}[1]{\bm{u_{#1}}}

\newcommand{\onlinever}[1]{#1}

\title{Nivat's conjecture holds for \\sums of two periodic configurations}
\author{Michal Szabados\thanks{Research supported by the Academy of Finland Grant 296018.}}
\institute{
	Department of Mathematics and Statistics,\\
	University of Turku, 20014 Turku, Finland
}

\begin{document}
\maketitle

\env{abstract}{
	Nivat's conjecture is a long-standing open combinatorial problem. It concerns two-dimensional configurations, that is, maps $\Z^2 \rightarrow \A$ where $\A$ is a finite set of symbols. Such configurations are often understood as colorings of a two-dimensional square grid. Let $P_c(m,n)$ denote the number of distinct $m \times n$ block patterns occurring in a configuration $c$. Configurations satisfying $P_c(m,n) \leq mn$ for some $m,n \in \N$ are said to have low rectangular complexity. Nivat conjectured that such configurations are necessarily periodic.
	
	Recently, Kari and the author showed that low complexity configurations can be decomposed into a sum of periodic configurations. In this paper we show that if there are at most two components, Nivat's conjecture holds. As a corollary we obtain an alternative proof of a result of Cyr and Kra: If there exist $m,n \in \N$ such that $P_c(m,n) \leq mn/2$, then $c$ is periodic. The technique used in this paper combines the algebraic approach of Kari and the author with balanced sets of Cyr and Kra.	
}


\section{Introduction}

Let $\A$ be a finite set of symbols and $d$ a positive integer, the dimension.  A \emph{$d$-dimensional symbolic configuration\/} $c$ is an element of $\A^{\Z^d}$, that is, a map assigning a symbol to every vertex of the lattice $\Z^d$. The symbol at position $\v \in \Z^d$ is denoted $c_\v$.

For a non-empty finite domain $D \subset \Z^d$, the elements of $\A^D$ are \emph{$D$-patterns}. We can observe patterns in a given configuration, the $D$-pattern occurring in $c$ at position $\v \in \Z^d$ is the map
\env{align*}{
	p \colon D &\rightarrow \A \\
	\u &\mapsto c_{\v+\u}.
}
The number of distinct $D$-patterns occurring in $c$, denoted $P_c(D)$, is the \emph{$D$-pattern complexity} of $c$. We say that $c$ has \emph{low complexity} if $P_c(D) \leq \abs D$ holds for some $D$.

We study what conditions on complexity imply that a configuration is periodic, that is, when there exists a non-zero vector $\u$ such that $c_\v = c_{\v+\u}$ for all $\v \in \Z^d$. The situation in one dimension was described by Morse and Hedlund \cite{MorseHedlund38}, let us denote $\llbracket n \rrbracket = \{0,\dots,n-1\}$:

\env{theorem*}{[Morse--Hedlund]
	Let $c$ be a one-dimensional symbolic configuration. Then $c$ is periodic if and only if there exists $n \in \N$ such that $P_c(\llbracket n \rrbracket) \leq n$.
}

As a corollary, non-periodic one-dimensional configurations satisfy $P_c(\llbracket n \rrbracket) \geq n+1$. Those for which equality holds for every $n$ are \emph{Sturmian words}, they are a central topic of combinatorics on words and have connections to discrete geometry, finite automata and mathematical physics \cite{Lothaire2002algebraic, Allouche2003automatic, Damanik1999}. Note that Sander and Tijdeman \cite{SanderTijdeman00} extended the Morse--Hedlund theorem for patterns of other shapes than $\llbracket n \rrbracket$, they showed that in fact any low complexity one-dimensional symbolic configuration is periodic.

Nivat's conjecture \cite{Nivat97} is a natural extension of the theorem to two-dimensions. To simplify notation we write $P_c(m,n) = P_c(\llbracket m \rrbracket \times \llbracket n \rrbracket)$.

\env{conjecture*}{[Nivat]
	If a two-dimensional symbolic configuration $c$ satisfies $P_c(m,n) \leq mn$ for some $m,n \in \N$, then it is periodic.
}

Nivat's conjecture is tight in the sense that there exist non-periodic configurations satisfying $P_c(m,n) = mn+1$ for all $m,n \in \N$, all such configurations were classified by Cassaigne \cite{Cassaigne99}. Note that the conjecture is not an equivalence, the opposite implication is easily seen to be false.

There have been a number of partial results towards the conjecture. Cyr and Kra \cite{CyrKra16} proved that having $P_c(3,n) \leq 3n$ for some $n \in \N$ implies periodicity, which was an improvement on a previous result with constant 2 \cite{SanderTijdeman02}. In another direction, there are results showing that having $P_c(m,n) \leq \alpha m n$ for some $m,n \in \N$ implies periodicity for a suitable real $\alpha$. The best result to date is also by Cyr and Kra \cite{CyrKra15} with $\alpha = 1/2$, which improved on previous constants $\alpha = 1/16$ \cite{QuasZamboni04} and $\alpha = 1/144$ \cite{EpifanioKoskasMignosi03}. Recently, Kari and the author \cite{KariSzabados2015ICALP} proved an asymptotic version of the conjecture: If $P_c(m,n) \leq mn$ for infinitely many pairs $(m,n) \in \N^2$, the configuration is periodic.

The Morse--Hedlund theorem does not analogously generalize to higher dimensions. There exists a three-dimensional configuration with low block complexity which is not periodic \cite{SanderTijdeman00}.

\subsection*{Our contributions}


In \cite{KariSzabados2015ICALP}, Kari and the author introduced an algebraic view on symbolic configurations. Following their definition, let a \emph{configuration} be any formal power series in $d$ variables $x_1, \dots, x_d$ with complex coefficients, that is, an element of
\env{align*}{
	\C[[X^{\pm1}]] = \makesetbig{\sum_{\v \in \Z^d} c_\v X^\v }{c_\v \in \C}
}
where $X^\v$ is a shorthand for $x_1^{v_1} \cdots x_d^{v_d}$.\footnote{For the most of this paper, however, it is enough to consider configurations to be elements of $\C^{\Z^d}$.} If the configuration has only integer coefficients it is called \emph{integral}, if they come from a finite set the configuration is \emph{finitary}. A symbolic configuration can be identified with a finitary integral configuration if the symbols from $\A$ are chosen to be integers. Kari and the author in \cite{KariSzabados2015ICALP} proved:

\env{theorem*}{[Decomposition theorem]
	Let $c$ be a low complexity $d$-dimensional finitary integral configuration. Then there exists $k \in \N$ and periodic $d$-dimensional configurations $c_1, \dots, c_k$ such that $c = c_1 + \dots + c_k$.
}

Note that the summands do not have to be finitary configurations. The minimal possible number of components $k$ in the decomposition plays an important role.
In this paper we prove:

\env{theorem}{
	\label{thm:main-thm}
	Let $c$ be a two-dimensional configuration satifying $P_c(m,n) \leq mn$ for some $m,n \in \N$. If $c$ is a sum of two periodic configurations then it is periodic.
}

In the proof of the asymptotic version of Nivat's conjecture given in \cite{KariSzabados2016Arxiv}, configurations which are a sum of horizontally and vertically periodic configuration had to be handled separately using a rather technical combinatorial approach. \autoref{thm:main-thm} is of particular interest since it covers this case.

In this paper we revisit the method of Van Cyr and Bryna Kra \cite{CyrKra15, CyrKra16}. They approach Nivat's conjecture from the point of view of symbolic dynamics. They use a refined version
of the classical notion of expansiveness of a subshift, a so called
\emph{one-sided non-expansiveness}. A key definition of theirs is that
of a \emph{balanced set} -- it is a shape $D \subset \Z^2$ which satisfies
a particular condition on the complexity $P_c(D)$.
(Note that this notion is different from balancedness usual in combinatorics on words.)
The crucial tool
they developed is a combinatorial lemma which links one-sided non-expansiveness
and balanced sets to periodicity of a configuration. However, in order to obtain
the main result of the paper from the lemma it still takes a rather lengthy
technical analysis.

We combine the algebraic method with ideas of Cyr and Kra. We start the exposition with a very basic introduction to the topic of symbolic dynamics. In \autoref{sec:symbolic-dynamics} we define a subshift, in \autoref{sec:geometry} we fix some geometric terminology, and in \autoref{sec:nonexpansiveness} we give definitions of non-expansiveness and one-sided non-expansiveness of a subshift.

In \autoref{sec:balanced-sets} we introduce a simplified version of a balanced set and prove \autoref{lem:cyr-kra} which connects balanced sets with periodicity using the ideas of Cyr and Kra. We use the lemma together with decomposition theorem to prove \autoref{thm:main-thm} in \autoref{sec:results}. As a corollary, we obtain an alternative proof of Theorem 1.2 of \cite{CyrKra15}, the main result of their paper:

\envparam[Cyr, Kra]{theorem*}{
	Let $c$ be a configuration satisfying $P_c(m,n) \leq mn/2$ for some
	$m,n \in \N$. Then $c$ is periodic.
}

\section{Symbolic dynamics and subshifts}
\label{sec:symbolic-dynamics}

Let us recall basic facts from symbolic dynamics, for a comprehensive reference and proofs see \cite{Kurka}.

Symbolic dynamics studies $\A^{\Z^d}$ as a topological space. Let us first make $\A$ a topological space by endowing it with the discrete topology. Then $\A^{\Z^d}$ is considered to be a topological space with the product topology.

Open sets in this topology are for example sets of the following form. Let $D \subset \Z^d$ be finite and $p \colon D \rightarrow \A$ arbitrary. Then
\env{align*}{
	Cyl(p) := \makesetbig{c \in \A^{\Z^d}}{\forall \v \in D \colon c_\v = p_\v}
}
is an open set, also called a \emph{cylinder}. In fact, the collection of cylinders $Cyl(p)$ for all possible $p$ forms a subbase of the topology on $\A^{\Z^d}$.

For a vector $\u \in \Z^d$, the \emph{shift} operator $\tau_\u \colon \A^{\Z^d} \rightarrow \A^{\Z^d}$ is defined by $(\tau_\u(c))_\v = c_{\v - \u}$. Informally, $\tau_\u$ shifts a configuration in the direction of vector $\u$.

The set $\A^{\Z^d}$ is called the \emph{full shift}. A subset $X \subset \A^{Z^d}$ is called a \emph{subshift} if it is a topologically closed set which is invariant under all shifts $\tau_\u$:
\env{align*}{
	\forall \u \in \Z^d \colon c \in X \Rightarrow \tau_\u(c) \in X.
}
Subshifts are the central objects of study in symbolic dynamics.

Let $c$ be a symbolic configuration. We denote by $X_c$ the \emph{orbit closure} of $c$, that is, the smallest subshift which contains $c$. It can be shown that $c$ contains exactly those configurations $c'$ whose finite patterns are among the finite patterns of $c$. In particular, for any $c' \in X_c$ and a finite domain $D$ we have $P_{c'}(D) \leq P_c(D)$.

\env{example}{
	\label{ex:cross_orbit_closure}
	Let us give an example of taking orbit closure. Let $c \in \{0,1\}^{\Z^2}$ be such that $c_{ij} = 1$ if $i = 0$ or $j = 0$, and $c_{ij} = 0$ otherwise. When pictured, the configuration $c$ consists of a large cross with its center at $(0,0)$. The orbit closure $X_c$ then consist of four types of configurations: a cross,  a horizontal line, a vertical line and all zero configurations, with all possible translations, see \autoref{fig:cross_orbit_closure}. It is easy to see that any pattern which occurs in them also occurs in $c$, and not difficult to prove that those are all such configurations.
\qed}

\env{figure}{
	\centering
	\includegraphics[scale=0.7]{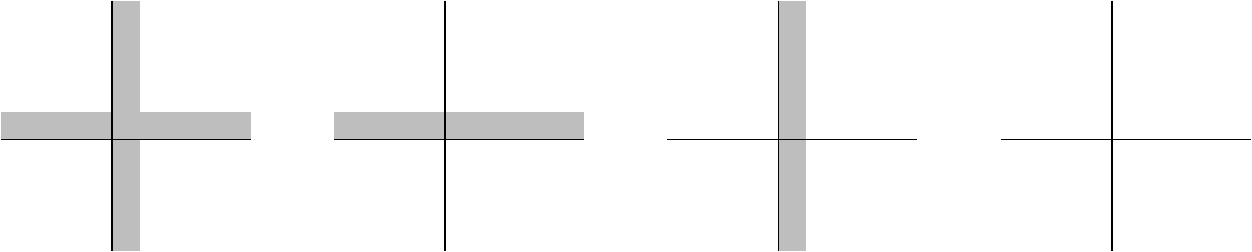}
	\caption{Four types of configurations in the orbit closure $X_c$ from \autoref{ex:cross_orbit_closure}. The gray color corresponds to value $1$, white is $0$.}
	\label{fig:cross_orbit_closure}
}

%

\section{Geometric notation and terminology}
\label{sec:geometry}

In the sequel we will be concerned with the geometry of $\Z^2$. Let us establish some notation and terminology.

We view $\Z^2$ as a subset of the vector space $\Q^2$. A \emph{direction} is an equivalence class of $\Q^2 \setminus \{(0,0)\}$ modulo the equivalence relation $u \sim v$ iff $u=\lambda v$ for some $\lambda>0$. By a slight abuse of notation, we identify a non-zero vector $\u \in \Z^2$ with the direction $\u \Q^+$.

Let $\u \in \Z^2$ be non-zero. An (undirected) \emph{line} in $\Z^2$ is a set of the form
$$\makeset{\v + q\u}{q \in \Q} \cap \Z^2$$
for some $\v \in \Z^2$. We call both $\u$ and $-\u$ a \emph{direction} of the line. We define a \emph{directed line} to be a line augmented with one of the two possible directions.

Let $\ell$ be a directed line in direction $\u$ going through $\v \in \Z^2$. The \emph{half-plane} determined by $\ell$ is defined by
\env{align*}{
	H_\ell = \makesetbig{\v + \w}{\w \in \Z^2, w_1 u_2 - u_1 w_2 \geq 0}.
}
With the usual choice of coordinates it is the half-plane ``on the right'' from the line.
Let $H_\u$ denote the half-plane determined by the directed line in direction $\u$ going through the origin.


We say that a non-empty $D \subset \Z^2$ is \emph{convex} if $D$ can be written as an intersection of half-planes. \emph{Convex hull} of $D$, denoted $Conv(D)$, is the smallest convex set containing $D$.  Assume $\ell$ is a directed line in direction $\u$ such that $D \subset H_\ell$ and $\ell \cap D$ is non-empty. If $\abs{\ell \cap D} > 1$ we call it the \emph{edge} of $D$ in direction $\u$, otherwise we call it the \emph{vertex} of $D$ in direction $\u$. Note that a vertex is a vertex for many directions, but an edge has a unique direction (as long as $D$ is not contained in a line). See \autoref{fig:convex_set} for an example.

\env{figure}{
	\centering
	\includegraphics[scale=0.6]{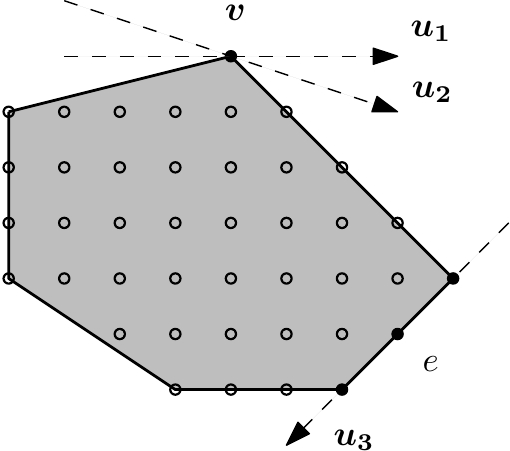}
	\caption{A convex set. The point $\v$ is a vertex of the set for both directions $\uu1$ and $\uu2$. The set of three marked points $e$ is the edge in direction $\uu3$.}
	\label{fig:convex_set}
}

Let $\u$ be a direction and $\ell, \ell'$ two directed lines in direction $\u$. If
$$S = H_\ell \setminus H_{\ell'}$$
is non-empty, then $S$ is called a \emph{stripe} in direction $\u$. We call $\ell, \ell'$ the \emph{inner} and \emph{outer} boundary of $S$ respectively. Let $S^\circ = S \setminus \ell$ be the \emph{interior} of $S$.

For $A, B \subset \Z^2$, we say that $A$ \emph{fits in} $B$ if there exists a translation of $A$ which is a subset of $B$.

\section{Non-expansiveness and one-sided non-expansiveness}
\label{sec:nonexpansiveness}

It can be verified that the topology on $\A^{\Z^d}$ is compact and also metrizable. Note that shift operators $\tau_\u$ are continuous maps on $\A^{\Z^d}$. Expansiveness can be defined in general for a continuous action on a compact metric space, the definition is however too general for our purposes. We give a definition specific to the case of $\A^{\Z^2}$.

Let $X \subset \A^{\Z^2}$ be a subshift and $\u$ a direction. Then $\u$ is an \emph{expansive direction} for $X$ if there exists a stripe $S$ in direction $\u$ such that
\env{align*}{
	\forall c, e \in X \colon \ \  c \restr S = e \restr S \ \   \Rightarrow \ \   c = e.
}
Informally speaking, $\u$ is an expansive direction for $X$ if a configuration in $X$ is uniquely determined by its coefficients in a wide enough stripe in direction $\u$.

A two-dimensional configuration is \emph{doubly periodic} if it has two linearly independent period vectors. The following classical theorem links double periodicity of a configuration with expansiveness. It is a corollary of a theorem by Boyle and Lind \cite{BoyleLind1997}.

\env{theorem}{
	Let $c$ be a symbolic configuration. Then $c$ is doubly periodic iff all directions are expansive for $X_c$.
\qed}


Let $X \subset \A^{\Z^2}$ be a subshift and $\u$ a direction. Then $\u$ is a \emph{one-sided expansive direction} for $X$ if
$$\forall c, e \in X \colon \ \  c \restr{H_\u} = e \restr{H_\u} \ \  \Rightarrow \ \  c = e.$$
Equivalently, $\u$ is a one-sided expansive direction for $X$ if there exists a wide enough stripe $S$ in direction $\u$ such that $\forall c, e \in X \colon c \restr S = e \restr S \Rightarrow c \restr{H_{-\u}} = e \restr{H_{-\u}}$. See \autoref{fig:expansiveness} for a comparison of the notion of expansiveness and one-sided expansiveness.

\env{figure}{
	\centering
	\includegraphics[scale=0.35]{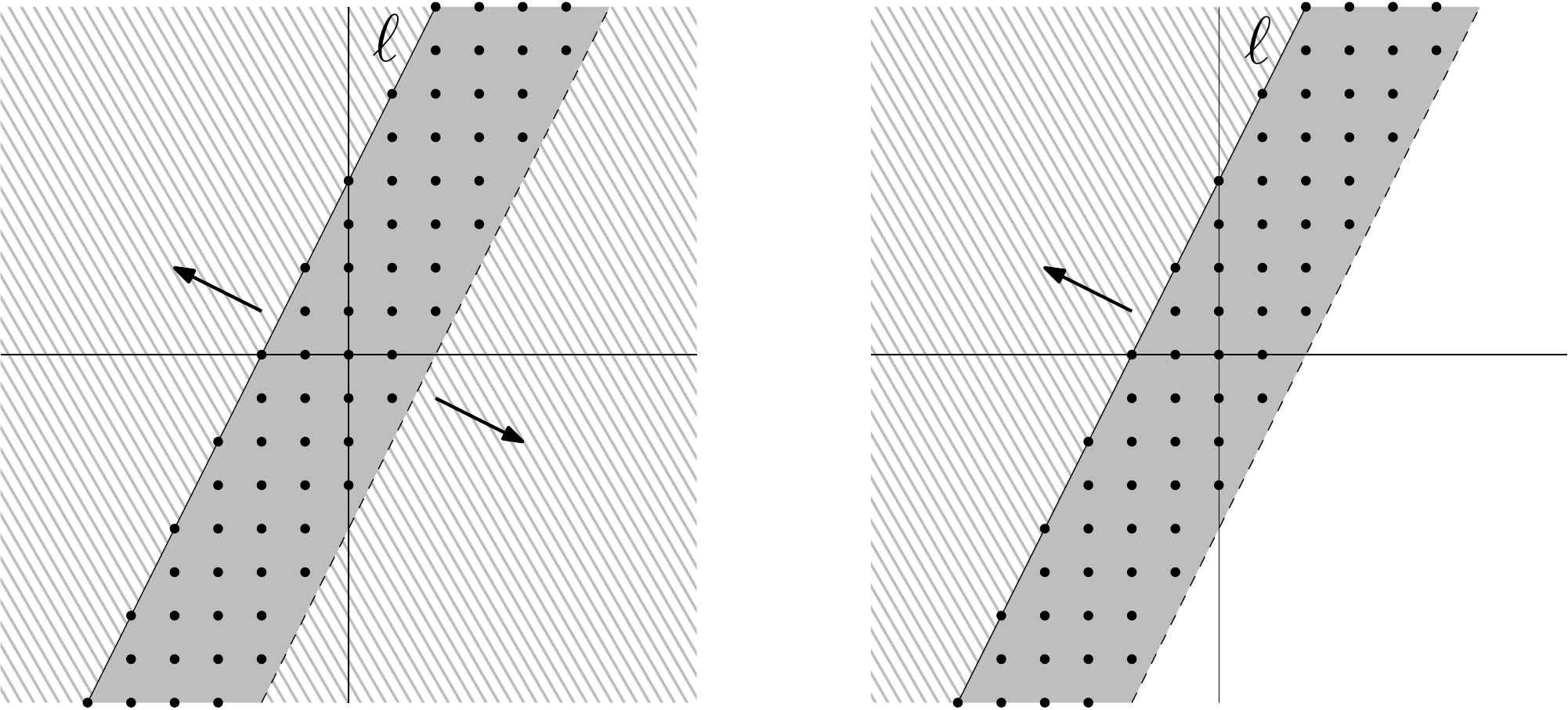}
	\caption{The figure on the left illustrates expansiveness -- values of the configuration inside the stripe determine the whole configuration. On the right we see one-sided expansiveness in direction $(1,2)$ -- values in the half-plane $H_\ell$, or equivalently in a wide enough stripe, determine the values in the half-plane $\Z^2 \setminus H_\ell$.}
	\label{fig:expansiveness}
}

\env{example}{[Ledrappier's subshift]
	It is possible for a subshift to be one-sided expansive but non-expansive in the same direction. Consider a subshift $X \subset \{0,1\}^{\Z^2}$ consisting of configurations $c$ which satisfy $c_{ij} \equiv c_{i,j+1} + c_{i+1,j+1} \pmod2$. Upper half-plane of a configuration determines the whole, since any single row determines the one below it. Therefore $(-1,0)$ is a one-sided expansive direction for $X$. However, no stripe in direction $(-1,0)$ determines a configuration from the subshift; for any row, there are always two possibilities for the row above it (they are complements of each other). Any horizontal stripe can be extended to the upper half-plane in infinitely many ways.
\qed}

We are primarily interested in non-expansive directions. In our setup, it is known that there are only finitely many of them, we omit the proof for space reasons. \onlinever{(See Appendix.)}

\env{lemma}{
	\label{lem:finitely-many-nonexp-dirs}
	Let $c$ be a low complexity two-dimensional configuration. Then there are at most finitely many one-sided non-expansive directions for $X_c$.
\qed}

For later use it will be practical to define non-expansiveness explicitly. Let $X \subset \A^{\Z^2}$ be a subshift and $S$ a stripe in direction $\u$. We say that $S$ is an \emph{ambiguous stripe in direction $\u$} if there exist $c, e \in X$ such that
\env{align}{
	c \restr{S^\circ} = e \restr{S^\circ},  \ \textrm{ but }\  c \restr S \ne e \restr S. \label{eq:ambig_stripe}
}
We say that $c \in X$ \emph{contains} an ambiguous stripe $S$ if there exists $e \in X$ satisfying (\ref{eq:ambig_stripe}). Informally, a stripe is ambiguous if its interior does not determine the inner boundary. 

\env{definition*}{
	Let $\u$ be a direction and $X \subset \A^{\Z^2}$ a subshift. Then $\u$ is \emph{one-sided non-expansive direction} if there exists an ambiguous stripe in direction $\u$ of arbitrary width.
}

We leave the proof that this is the converse of the earlier definition of one-sided expansiveness to the reader.

\section{Balanced sets}
\label{sec:balanced-sets}

Let $c$ be a fixed symbolic configuration.

\env{definition}{
	\label{def:balanced-set}
	Let $B \subset \Z^2$ be a finite and convex set, $\u$ a direction and $E$ an edge or a vertex of $B$ in direction $\u$. Then $B$ is $\u$-balanced if:
	\envparam[(i)]{enumerate}{
		\item $P_c(B) \leq \abs B$
		\item $P_c(B) <  P_c(B \setminus E) + \abs E$ \label{few_extensions}
		\item Intersection of $B$ with all lines in direction $\u$ is either empty or of size at least $\abs E - 1$. \label{long_cuts}
	}
}

The three conditions of the definition can be interpreted as follows. The first one simply states that $B$ is a low complexity shape. The second condition limits the number of $(B \setminus E)$-patterns which do not extend uniquely to a $B$-pattern, there is strictly less than $\abs E$ of them. The third condition is implied if the length of the edge in direction $\u$ is smaller or equal to the length of the edge in the opposite direction, as can be seen in the next proof.

\env{lemma}{
	\label{lem:balanced-set-existence}
	Let $c$ be such that $P_c(m,n) \leq mn$ holds for some $m,n \in \N$ and $\u$ be a direction. Then there exists a $\u$-balanced or $(-\u)$-balanced set. Moreover, if $\u$ is horizontal or vertical, then there exists a $\u$-balanced set. 
}
\env{proof}{
	Let $D$ be an $m \times n$ rectangle, we have $P_c(D) \leq \abs D$. Let us define a sequence of convex shapes $D = D_0 \supset D_1 \supset \dots \supset D_k = \emptyset$ such that $D_{i} \setminus D_{i+1}$ is the edge of $D_i$ in direction $(-1)^i\u$. Informally, the sequence represents shaving off an edge (or a vertex) of the shape alternately in directions $\u$ and $-\u$. See \autoref{fig:balanced_set_existence} for an illustration.
	
	Consider the expression $P_c(D_i) - \abs{D_i}$ as a function of $i$. For $i = 0$ its value is non-positive and for $i = k$ its value is $1$. Let $i \in [0,k-1]$ be smallest such that $0 < P_c(D_{i+1}) - \abs{D_{i+1}}$, then we have 
	$$P_c(D_i) - \abs{D_i} \leq 0 < P_c(D_{i+1}) - \abs{D_{i+1}}.$$
	Denote $E = D_i \setminus D_{i+1}$, it is an edge or a vertex of $D_i$ in direction $\u$ or $-\u$. Adding $\abs{D_i}$ to the inequality and rewriting gives $P(D_i) \leq \abs{D_i} < P(D_i \setminus E) + \abs E$.
	
	We show that $B = D_i$ is a balanced set by showing that \textit{(\ref{long_cuts})} of \autoref{def:balanced-set} holds. Without loss of generality let the direction of $E$ be $\u$. Then, by construction, the length of $E$ is smaller or equal to the edge in direction $-\u$. In fact, if we consider the convex hull of $B$ in $\Q^2$, any line in direction $\u$ intersects it in a line segment longer or equal to $d$, the length of the edge. Any line segment of length at least $d$ in direction $\u$ intersects either none or at least $\abs E - 1$ integer points, and we are done.
	
	If $\u$ is either horizontal or vertical, instead of alternating the direction of shaved off edges, we can always shave off the edge in direction $\u$. It will be always the shortest edge in direction $\u$, therefore verification of part \textit{(\ref{long_cuts})} goes through.
\qed}

\env{figure}{
	\centering
	\includegraphics[scale=1.0]{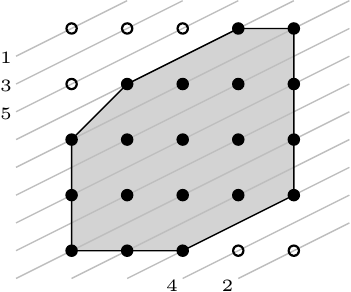}
	\caption{Shaving off edges or vertices of a $5 \times 5$ rectangle alternately in directions $(2, 1)$ and $(-2, -1)$. Small numbers indicate the order in which the edges or vertices were removed.}
	\label{fig:balanced_set_existence}
}

Next we present \autoref{lem:cyr-kra} which connects non-expansiveness and balanced sets with periodicity, based on the method of Cyr and Kra. Periodicity in the proof first arises in a stripe from the use of Morse--Hedlund theorem. This part of the proof follows Lemma 2.24 from \cite{CyrKra15}. The periodicity is then extended to the whole configuration by the following lemma, which is a corollary of Lemma 39 from \cite{KariSzabados2016Arxiv}. We omit the proof for space reasons. \onlinever{(See Appendix.)}

\env{lemma}{
	\label{lem:periodic-stripe-extends}
	Let $c$ be a two-dimensional configuration and $D$ a non-empty finite subset of $\Z^2$ such that $P_c(D) \leq \abs D$. Let $S$ be a stripe in direction $\u$ such that $D$ fits in $S$. If $S^\circ$ is periodic with a period in direction $\u$ then also $c$ is periodic with a period in direction $\u$.
\qed}

\begin{figure}
	\centering
	\begin{subfigure}{.5\textwidth}
		\centering
		\includegraphics[scale=0.4]{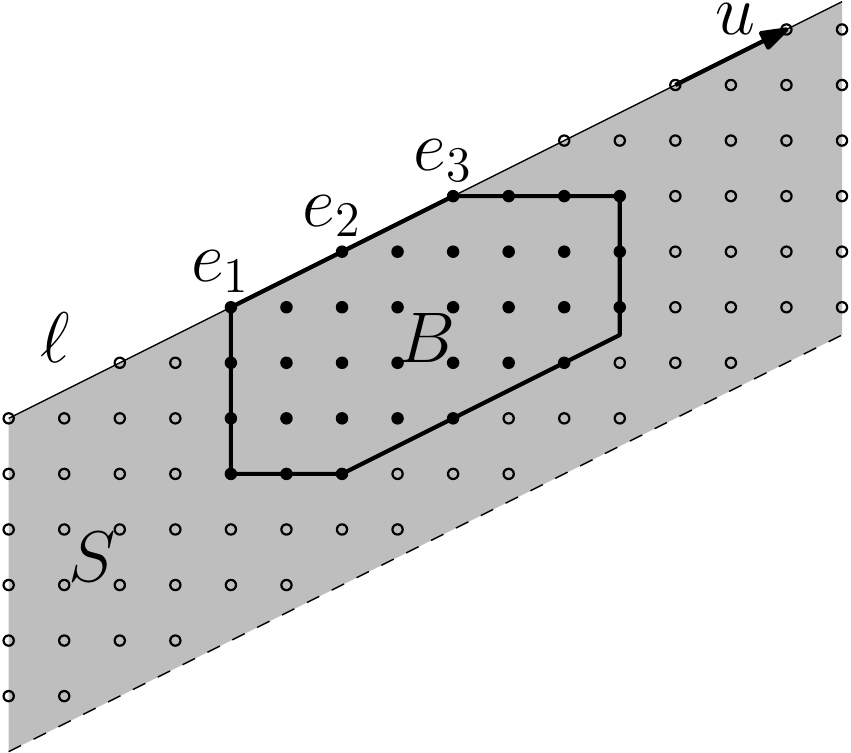}
	\end{subfigure}%
	\begin{subfigure}{.5\textwidth}
		\centering
		\includegraphics[scale=0.4]{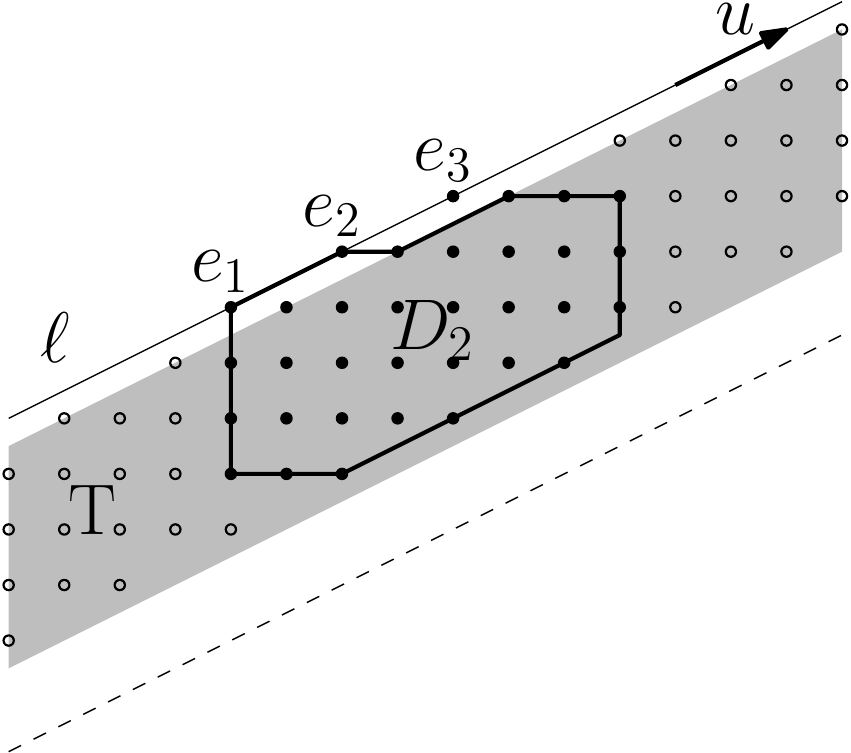}
	\end{subfigure}
	\caption{Illustration of the proof of \autoref{lem:cyr-kra}.}
	\label{fig:lemma_cyr_kra}
\end{figure}

\env{lemma}{
	\label{lem:cyr-kra}
	Let $c$ be a configuration and $B$ a $\u$-balanced set. Assume that $c$ contains an ambiguous stripe for $X_c$ in direction $\u$ such that $B$ fits in the stripe. Then $c$ is periodic in direction $\u$.
}
\env{proof}{
	Let $E$ be the edge or vertex of $B$ in direction $\u$, denote $S$ the stripe and let $\ell$ be the inner boundary of $S$ in direction $\u$. Without loss of generality assume $B \subset S$, $E \subset \ell$, and that $\u$ is not an integer multiple of a smaller vector. Let $e \in X_c$ be such that \autoref{eq:ambig_stripe} holds.
	
	Denote points in $E$ consecutively by $e_1, \dots, e_n$ (see \autoref{fig:lemma_cyr_kra}). Define a sequence $B = D_n \supset \dots \supset D_1 \supset D_0 = B \setminus E$ by setting  $D_{i-1} = D_i \setminus \{e_i\}$. Consider the values $P(D_i) - \abs{D_i}$. Since $B$ is a balanced set, by \textit{(\ref{few_extensions})} we have $P_c(D_n) - \abs{D_n} < P_c(D_0) - \abs{D_0}$, let $k \in [0,n-1]$ be such that 
	$$P_c(D_{k+1}) - \abs{D_{k+1}} < P_c(D_k) - \abs{D_k}.$$
	Adding $\abs{D_{k+1}}$ to both sides yields $P_c(D_{k+1}) < P_c(D_k) + 1$. On the other hand, $P_c(D_k) \leq P_c(D_{k+1})$ since $D_k \subset D_{k+1}$, and therefore we have $P_c(D_k) = P(D_{k+1})$. In other words, a $D_k$-pattern uniquely determines the value at position $e_{k+1}$.
	
	We will show that $\forall i \colon c \restr{D_k+i\u}\, \ne e \restr{D_k+i\u}$. For the contrary, assume that there is $j$ such that $c \restr{D_k+j\u} = e \restr{D_k+j\u}$. Using the property of $D_k$, we have $c \restr{e_{k+1}+j\u} = e \restr{e_{k+1}+j\u}$. Therefore $c \restr{D_k + (j+1) \u} = e \restr{D_k + (j+1) \u}$ and we can proceed by induction to show $c \restr{D_k + j' \u} = e \restr{D_k + j' \u}$ for all $j' > j$. Analogously, by constructing sets $D_i$ by removing edge points from the other end, it can be shown that also $c \restr{D_k + j' \u} = e \restr{D_k + j' \u}$ for all $j' < j$. We proved $c \restr S = e \restr S$, which is a contradiction with ambiguity of $S$.
	
	We have that all $(B \setminus E)$-patterns $c \restr{(B \setminus E) + i \u}$ have at least two possible extensions into a $B$-pattern. Part \textit{(\ref{few_extensions})} of \autoref{def:balanced-set} implies that there are at most $\abs E - 1$ such patterns. Let $T$ be a thinner stripe in direction $\u$ defined by $T = \bigcup_{i \in \Z} (B \setminus E) + i \u$. Using part \textit{(\ref{long_cuts})} of \autoref{def:balanced-set}, values of $c$ on every line $\lambda \subset T$ in direction $\u$ contain at most $\abs E - 1$ distinct subsegments of length at least $\abs E - 1$. By Morse--Hedlund theorem, the values on the line repeat periodically. Therefore $c \restr{T}$ is periodic in direction $\u$.
	
	$B$ fits in the stripe $T \cup \ell$ and its interior $T$ is periodic in direction $\u$. By \autoref{lem:periodic-stripe-extends} also $c$ is periodic in direction $\u$.
\qed}

\section{Main result}
\label{sec:results}

\env{theorem*}{[\autoref{thm:main-thm}]
	Let $c$ be a two-dimensional configuration satisfying $P_c(m,n) \leq mn$ for some $m,n \in \N$. If $c$ is a sum of two periodic configurations then it is periodic.
}
\env{proof}{
	For contradiction assume $c$ is non-periodic and denote $c_1, c_2$ periodic configurations such that $c = c_1 + c_2$. Let $\uu1, \uu2$ be their respective vectors of periodicity. If they are linearly dependent, $c$ is periodic and we are done. Otherwise, define a parallelogram
	\env{align*}{
		D = \makesetbig{a \uu1 + b \uu2}{a, b \in [0, 1)} \cap \Z^2.
	}
	We can choose $\uu1, \uu2$ large enough so that an $m \times n$ rectangle fits in. We can also assume that $\uu2 \in H_{\uu1}$. Denote $D_j = D + j \uu2$ and define a sequence of stripes $S_j = \bigcup_{i \in \Z} D_j + i\uu1$. The setup is illustrated in \autoref{fig:nivat_ord2_1}.
	
	Assume that there are $j \ne j'$ such that $c \restr{D_j} = c \restr{D_{j'}}$. We claim that then $c \restr{S_j} = c \restr{S_{j'}}$. Note that since $c = c_1 + c_2$, for $\v \in \Z^2$ we have
	\env{align*}{
		(c_{(\v+\uu1)+j\uu2} - c_{(\v+\uu1)+j'\uu2}) - (c_{\v+j\uu2} - c_{\v+j'\uu2}) = 0.
	}
	In particular, if $c_{\v+j\uu2} = c_{\v+j'\uu2}$, then also $c_{(\v+\uu1)+j\uu2} = c_{(\v+\uu1)+j'\uu2}$. Since $c_{\v+j\uu2} = c_{\v+j'\uu2}$ holds for $\v \in D$, it also holds for $\v \in D + \uu1$, and by induction $c \restr{S_j} = c \restr{S_{j'}}$.

	Since $c$ is finitary there are only finitely many possible $D$-patterns, let $N$ be an upper bound on their number. There are also finitely many stripe patterns $c \restr{S_j}$ since the pattern in $S_j$ is determined by the pattern in $D_{j}$. Because $c$ is not periodic, there exists $k \in \Z$ such that $c \restr{S_k} \ne c \restr {S_{k - N!}}$.
	
	By \autoref{lem:balanced-set-existence}, there is either a $\uu1$-balanced or $(-\uu1)$-balanced set $B$, without loss of generality assume the former. Since $c$ is non-periodic, by \autoref{lem:cyr-kra} there is no ambiguous stripe in $c$ in direction $\uu1$ in which $B$ fits. $B$ fits in any stripe $S_j$, therefore values in any stripe $S_j$ determine the values in the whole half-plane on the side of the inner boundary of $S_j$.
	
	By pigeonhole principle, there are $j < j' \in [0, N]$ such that $c \restr {S_{k+j}} = c \restr {S_{k+j'}}$. The two stripes extend uniquely to the half-planes on the side of their inner boundary. Therefore the half-plane $H = \bigcup_{i \leq j'} S_i$ has period $(j'-j)\uu2$. Since $j'-j$ divides $N!$ and $S_k, S_{k-N!} \subset H$, we have a contradiction with $c \restr{S_k} \ne c \restr {S_{k - N!}}$.
\qed}

\env{figure}{
	\centering
	\includegraphics[scale=0.35]{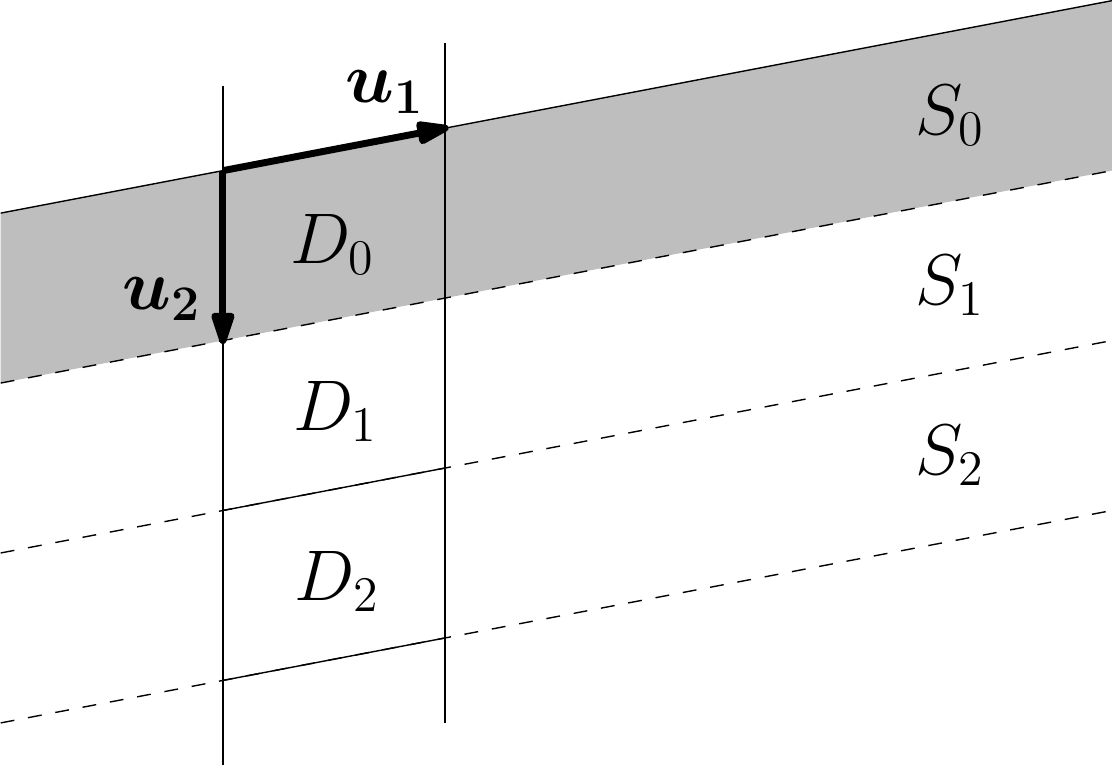}
	\caption{Proof of \autoref{thm:main-thm}.}
	\label{fig:nivat_ord2_1}
}

\env{corollary}{
	If a non-periodic configuration $c$ is a sum of two periodic ones, then $P_c(m,n) \geq mn+1$ for all $m,n \in \N$.
\qed}

We finish the exposition by reproving the result of Cyr and Kra from \cite{CyrKra15}. To do that, we need additional theory from \cite{KariSzabados2016Arxiv}. Multiplication of a two-dimensional configuration $c$ by a polynomial $f \in \C[x_1, x_2]$ is well defined. If $fc = 0$, we call $f$ an \emph{annihilator} of $c$.
The following two lemmas we state without a proof, they are direct corollaries of Corollary 24 and Lemma 32 of \cite{KariSzabados2016Arxiv}, respectively.

\env{lemma}{
	\label{lem:explicit-decomposition}
	Let $c$ be a low complexity two-dimensional integral configuration. Then there exists $k \in \N$ and polynomials $\phi_1, \dots, \phi_k \in \C[x_1,x_2]$ with the following properties:
	 
	Every annihilator of $c$ is divisible by $\phi_1 \cdots \phi_k$. Furthermore, $c$ can be written as a sum of $k$, but no fewer periodic configurations. If $g$ is a product of $0 \leq \ell < k$ of the polynomials $\phi_i$, then $gc$ can be written as a sum of $k-\ell$, but no fewer periodic configurations.
\qed}

Any polynomial in $\C[x_1, x_2]$ can be written as $f = \sum_{\v \in \Z^2} a_\v X^\v$. The \emph{support} of $f$, denoted $\supp(f)$, is defined as the finite set of vectors $\v \in \Z^2$ such that $a_\v \ne 0$. We say that $f$ \emph{fits in} a subset $D \subset \Z^2$ if its support fits in $D$.

\env{lemma}{
	\label{lem:normalized}
	Let $c$ be a finitary configuration. Then the symbols of $\A$ can be changed to suitable integers such that if $P_c(D) \leq \abs D$ for some $D \subset \Z^d$, then there exists an annihilator $f$ which fits in $-D$.
\qed}

\env{theorem}{
	\label{thm:nivat-mn-over-2}
	Let $c$ be a configuration such that $P_c(m,n) \leq mn/2$ for some $m,n \in \N$. Then $c$ is periodic.
}
\env{proof}{
	Assume that the symbols of $\A$ have been renamed as in \autoref{lem:normalized}, then there exists $f$ an annihilator of $c$ which fits in an $m \times n$ rectangle. By \autoref{lem:explicit-decomposition}, we can write $f = \phi_1 \cdots \phi_k h$. If $k \leq 2$ then $c$ is periodic by \autoref{thm:main-thm}. Assume $k \geq 3$, we will show that it leads to a contradiction.
	
	Let $g = \phi_3 \cdots \phi_k$,  $c' = gc$ and let $m_g, n_g \in \N$ be smallest such that $g$ fits in an $(m_g+1)\times(n_g+1)$ rectangle, see \autoref{fig:mn_over_2}. Note that an $(m - m_g)\times(n - n_g)$ block  in $c'$ is determined by multiplication by $g$ from an $m \times n$ block in $c$. Therefore $P_c(m,n) \geq P_{c'}(m-m_g, n-n_g)$.
	
	By \autoref{lem:explicit-decomposition}, $c'$ is a sum of two but no fewer periodic configurations. Thus it is not periodic, and by \autoref{thm:main-thm},
	$$P_c(m,n) \geq P_{c'}(m-m_g, n-n_g) > (m-m_g)(n-n_g).$$
	
	Let $\v$ be an arbitrary vertex of the convex hull of $-\supp(g)$. Consider all translations of $-\supp(g)$ which are a subset of the rectangle $\llbracket m \rrbracket \times \llbracket n \rrbracket$, denote $R$ the locus of $\v$ under these translations. There are $(m-m_g)(n-n_g)$ such translations, therefore the size of $R$ is the same number.
	
	Now let us define a shape $U = \llbracket m \rrbracket \times \llbracket n \rrbracket \setminus R$. It is a shape such that no polynomial multiple of $g$ fits in $-U$. In particular no annihilator of $c$ fits in $-U$, and thus by \autoref{lem:normalized},
	\env{align*}{
		P_c(m,n) \geq P_c(U) > \abs U.
	}
	Since either $(m-m_g)(n-n_g) = \abs R \geq mn/2$ or $\abs U \geq mn/2$, we have $P_c(m,n) > mn/2$, a contradiction.
\qed}

\env{figure}{
	\centering
	\includegraphics[scale=0.8]{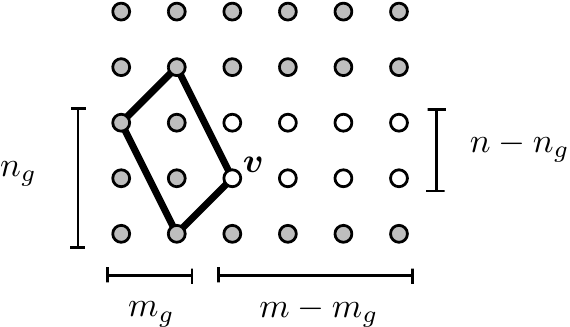}
	\caption{Proof of \autoref{thm:nivat-mn-over-2}. The quadrilateral depicts the convex hull of $-\supp(g)$ for a polynomial $g$, positioned in the bottom left corner of an $m \times n$ block. The white points form the set $R$ and the shaded points form the set $U$. We have $\abs U \geq mn/2$ or $\abs R \geq mn/2$.}
	\label{fig:mn_over_2}
}


	\newpage
	\bibliographystyle{alpha}
	\bibliography{bibliography}

\begin{thebibliography}{EKM03}

\bibitem[AS03]{Allouche2003automatic}
J.P. Allouche and J.~Shallit.
\newblock {\em Automatic Sequences: Theory, Applications, Generalizations}.
\newblock Cambridge University Press, 2003.

\bibitem[BL97]{BoyleLind1997}
Mike Boyle and Douglas Lind.
\newblock Expansive subdynamics.
\newblock {\em Transactions of the American Mathematical Society},
  349(1):55--102, 1997.

\bibitem[Cas99]{Cassaigne99}
Julien Cassaigne.
\newblock Double sequences with complexity mn+1.
\newblock {\em Journal of Automata, Languages and Combinatorics},
  4(3):153--170, 1999.

\bibitem[CK15]{CyrKra15}
Van Cyr and Bryna Kra.
\newblock Nonexpansive {$\Bbb{Z}^2$}-subdynamics and {N}ivat's conjecture.
\newblock {\em Trans. Amer. Math. Soc.}, 367(9):6487--6537, 2015.

\bibitem[CK16]{CyrKra16}
Van Cyr and Bryna Kra.
\newblock Complexity of short rectangles and periodicity.
\newblock {\em European Journal of Combinatorics}, 52, Part A:146 -- 173, 2016.

\bibitem[DL99]{Damanik1999}
David Damanik and Daniel Lenz.
\newblock Uniform spectral properties of one-dimensional quasicrystals, i.
  absence of eigenvalues.
\newblock {\em Communications in Mathematical Physics}, 207(3):687--696, 1999.

\bibitem[EKM03]{EpifanioKoskasMignosi03}
Chiara Epifanio, Michel Koskas, and Filippo Mignosi.
\newblock On a conjecture on bidimensional words.
\newblock {\em Theor. Comput. Sci.}, 1-3(299), 2003.

\bibitem[KS15]{KariSzabados2015ICALP}
Jarkko Kari and Michal Szabados.
\newblock An algebraic geometric approach to {Nivat's} conjecture.
\newblock In {\em Automata, Languages, and Programming - 42nd International
  Colloquium, {ICALP} 2015, Kyoto, Japan, July 6-10, 2015, Proceedings, Part
  {II}}, pages 273--285, 2015.

\bibitem[KS16]{KariSzabados2016Arxiv}
Jarkko Kari and Michal Szabados.
\newblock An algebraic geometric approach to {Nivat's} conjecture.
\newblock \texttt{arXiv:1605.05929}, 2016.

\bibitem[K{\r u}r03]{Kurka}
P.~K{\r u}rka.
\newblock {\em Topological and Symbolic Dynamics}.
\newblock Collection SMF. Soci{\'e}t{\'e} math{\'e}matique de France, 2003.

\bibitem[Lot02]{Lothaire2002algebraic}
M.~Lothaire.
\newblock {\em Algebraic Combinatorics on Words}.
\newblock Encyclopedia of Mathematics an. Cambridge University Press, 2002.

\bibitem[MH38]{MorseHedlund38}
Marston Morse and Gustav~A. Hedlund.
\newblock Symbolic dynamics.
\newblock {\em American Journal of Mathematics}, 60(4):pp. 815--866, 1938.

\bibitem[Niv97]{Nivat97}
M.~Nivat.
\newblock Invited talk at {ICALP}, {Bologna}, 1997.

\bibitem[QZ04]{QuasZamboni04}
Anthony Quas and Luca~Q. Zamboni.
\newblock Periodicity and local complexity.
\newblock {\em Theor. Comput. Sci.}, 319(1-3):229--240, 2004.

\bibitem[ST00]{SanderTijdeman00}
J.~W. Sander and Robert Tijdeman.
\newblock The complexity of functions on lattices.
\newblock {\em Theor. Comput. Sci.}, 246(1-2):195--225, 2000.

\bibitem[ST02]{SanderTijdeman02}
J.~W. Sander and Robert Tijdeman.
\newblock The rectangle complexity of functions on two-dimensional lattices.
\newblock {\em Theor. Comput. Sci.}, 270(1-2):857--863, 2002.

\end{thebibliography}
	

	
	\onlinever{
		\newpage

\appendix

\section{Appendix}
\label{appendix}

Proofs in the appendix use definitions from \autoref{sec:results}.

\subsection{Proof of \autoref{lem:finitely-many-nonexp-dirs}}

The lemma also follows from existence of generating sets introduced by Cyr and Kra \cite{CyrKra15}. Here we show a proof using polynomials:

\env{proof}{
	By Lemma~5 of \cite{KariSzabados2016Arxiv}, there exists a non-trivial annihilator of the configuration. Let $F$ denote convex hull of it support. It has finitely many edges. We claim that only directions of the edges can be one-sided non-expansive for $X_c$.
	
	Let $\u$ be a direction such that $F$ has a vertex $\v$ in direction $\u$. Let $\ell$ be the line in direction $\u$ which is the closest to $H_\u$ but lies outside of $H_\u$. Then $F$ can be translated such that $F \setminus \{\v\}$ lies in $H_\u$ and $\v \in \ell$. Linear combination given by the annihilator determines the value of $c_\v$ from values in $H_\u$, and by translation in the whole line $\ell$. Moving to the next and next line in direction $\u$, all the values of $c$ are determined. We proved that $\u$ is a one-sided expansive direction for $X_c$.
\qed}

\subsection{Proof of \autoref{lem:periodic-stripe-extends}}

%

The proof is by reduction to Lemma 39 of \cite{KariSzabados2016Arxiv}:

\env{lemma*}{[Lemma~39]
	Let $c$ be a counterexample candidate and $\v \in \Z^2$ a non-zero vector. Let $S$ be an infinite stripe in the direction of $\v$ of maximal width such that $\phi$ does not fit in. Then $c$ restricted to the stripe $S$ is non-periodic in the direction of $\v$.
}

We assume the reader is comfortable with notions used in its statement. Let us however briefly describe some of them. A two-dimensional configuration is a \emph{counterexample candidate} if it is normalized non-periodic finitary integral configuration which has a non-trivial annihilator. Without going into further details, \emph{normalized} configurations have the property from \autoref{lem:normalized} and any configuration can be made normalized by changing the symbols in $\A$. The polynomial $\phi$ is the largest polynomial (w.r.t. polynomial division) which divides every annihilator, it is product of polynomials $\phi_i$ from the statement of \autoref{lem:explicit-decomposition}.

\env{proof}{[of \autoref{lem:periodic-stripe-extends}]
	Without loss of generality assume that $c$ is normalized and for the contrary assume that it is non-periodic, then $c$ is a counterexample candidate. By \autoref{lem:normalized} there is an annihilator which fits in $-D$ and therefore also in $S$. Then also $\phi$ fits in $S$. Let $T \subset S^\circ$ be a stripe in direction $\u$ of maximal width such that $\phi$ does not fit in. Since $c \restr T$ is periodic in direction $\u$, by Lemma~39 also $c$ is.
\qed}

	}

\end{document}